\documentclass{aa}

\usepackage{amsmath}
\usepackage{amssymb}
\usepackage{txfonts}
\usepackage{epsfig,graphicx}
\usepackage{xcolor}
\usepackage[colorlinks=true, linkcolor=blue, citecolor=blue, urlcolor=blue]{hyperref}
\usepackage{natbib}
\bibpunct{(}{)}{;}{a}{}{,} 
\newcommand{\txd}{{\text{d}}}

\newcommand{\bfI}{{\boldsymbol{I}}}
\newcommand{\bfK}{{\boldsymbol{K}}}

\newcommand{\calX}{{\mathcal{X}}}

\begin{document}

\title{Optical depth in polarised Monte Carlo radiative transfer}
\titlerunning{Optical depth in polarised Monte Carlo radiative transfer}

\author{
Maarten Baes\inst{\ref{UGent}} 
\and 
Christian Peest\inst{\ref{UGent},\ref{ESO}}
\and
Peter Camps\inst{\ref{UGent}}
\and
Ralf Siebenmorgen\inst{\ref{ESO}} 
}

\authorrunning{M. Baes et al.}

\institute{
Sterrenkundig Observatorium, Universiteit Gent, Krijgslaan 281 S9, 9000 Gent, Belgium
\label{UGent}
\and
European Southern Observatory, Karl-Schwarzschild-Stra{\ss}e 2, 85748 Garching bei M\"{u}nchen, Germany
\label{ESO}
}

\abstract{The Monte Carlo method is the most widely used method to solve radiative transfer problems in astronomy, especially in a fully general 3D geometry. A crucial concept in any Monte Carlo radiative transfer code is the random generation of the next interaction location. In polarised Monte Carlo radiative transfer with aligned non-spherical grains, the nature of dichroism complicates the concept of optical depth.}%
{We investigate in detail the relation between optical depth and the optical properties and density of the attenuating medium in polarised Monte Carlo radiative transfer codes that take into account dichroic extinction.}%
{Based on solutions for the radiative transfer equation, we discuss the optical depth scale in polarised radiative transfer with spheroidal grains. We compare the dichroic optical depth to the extinction and total optical depth scale.}%
{In a dichroic medium, the optical depth is not equal to the usual extinction optical depth, nor to the total optical depth. For representative values of the optical properties of dust grains, the dichroic optical depth can differ from the extinction or total optical depth by several ten percent. A closed expression for the dichroic optical depth cannot be given, but it can be derived efficiently through an algorithm that is based on the analytical result corresponding to elongated grains with a uniform grain alignment.}%
{Optical depth is more complex in dichroic media than in systems without dichroic attenuation, and this complexity needs to be considered when generating random free path lengths in Monte Carlo radiative transfer simulations. There is no benefit in using approximations instead of the dichroic optical depth.}


\maketitle

\section{Introduction}

Radiative transfer is a broad field in astronomy and beyond that aims to describe the interaction between radiation and matter. In an astronomical context, the Monte Carlo method is by far the most widely used method to solve radiative transfer problems. In the past decades, many different Monte Carlo codes have been developed to address different astrophysical radiative transfer problems, including photoionisation, absorption and scattering by cosmic dust, the origin of infrared emission, and resonant line scattering heating \citep[e.g.,][]{2001ApJ...551..269G, 2003MNRAS.340.1136E, 2011A&A...536A..79R, 2012MNRAS.424..884Y, 2013ApJS..207...30W, 2015A&C.....9...20C, 2016A&A...593A..87R}. General reviews on Monte Carlo transport can be found in e.g.\ \citet{DupreeFraley} or \citet{2008mcm..book.....K}, and dedicated reviews on radiative transfer in astrophysics include \citet{2011BASI...39..101W} and \citet{2013ARA&A..51...63S}. 

The essence of the Monte Carlo method is to represent the radiation field as the flow of a large but finite number of photon packages. The life cycle of each photon package is followed individually, and at every stage in this life cycle, the characteristics that determine the path of each photon package are determined in a probabilistic way by generating random numbers from the appropriate probability density function (PDF). At the end of the simulation, the radiation field, or more specifically the intensity of the radiation field, is recovered from a statistical analysis of the photon package paths. 

An important ingredient of Monte Carlo radiative transfer is the knowledge of the appropriate PDF for a given characteristic, and the accurate and efficient sampling of random numbers from these PDFs. For some characteristics, the PDFs are simple and sampling random numbers from them is trivial. For example, most sources send out radiation isotropically, which implies that the generation of propagation directions after an emission event simply comes down to generating a random point on the unit sphere. The PDF that controls the random starting positions for the photon package is dictated by the 3D luminosity density of the sources, and specific methods have been developed to generate such random positions from a range of 3D density distributions \citep{2015A&C....12...33B}.

An aspect that is central to any Monte Carlo radiative transfer code is the random generation of the next interaction location. More specifically, if a photon package is emitted or scattered into a given direction, one needs to randomly generate a free path length $s$ to the next interaction. To do so, we need to know the appropriate PDF $p(s)$. In this context, the concept of optical depth $\tau$ plays a crucial role. In optical depth space, the PDF $p(\tau)$ is a simple exponential distribution \citep{Cashwell1959, 2013ARA&A..51...63S}. This implies that the next interaction location can be found by randomly generating a random optical depth from an exponential distribution, and converting this optical depth to a physical path length. This last aspect is a critical point: we need to know the relation $\tau(s)$, or inversely $s(\tau)$ to properly calculate the next interaction location. 

In radiative transfer problems where polarisation is not taken into account, $\tau(s)$ can immediately be calculated from the density and the optical properties along the path, and does not depend on the intensity of the radiation field. The situation becomes more complex when polarisation comes into play. In the case of spherical or randomly oriented particles, polarised Monte Carlo radiative transfer is only slightly more difficult than unpolarised Monte Carlo radiative transfer. The main added complexity is that a Stokes vector needs to be introduced to characterise the polarisation status of each photon package, and that this Stokes vector can alter during a scattering event \citep[e.g.,][]{1994A&A...284..187F, 1995ApJ...441..400C, 1996ApJ...465..127B, 2017A&A...601A..92P}. The relation between optical depth and path length is the same as for unpolarised Monte Carlo radiative transfer.

The real complexity arises when the attenuating particles are non-spherical and aligned, as in the case of elongated dust grains in the interstellar medium. Such dust grains will be aligned with respect to the magnetic fields through a variety of processes \citep{1951ApJ...114..206D, 1967ApJ...147..943J, 1983ApJ...272..551A, 1994MNRAS.268..713L, 2007JQSRT.106..225L}. An interesting feature in this context is dichroism, which means that radiation of different polarisation experiences different amounts of extinction. The nature of dichroism complicates the relation between optical depth and the physical path length. 

In this paper, we investigate in detail whether an optical depth scale can still be used in a meaningful way in polarised Monte Carlo radiative transfer codes that take into account dichroic extinction. In Sect.~{\ref{Unpolarised.sec}} we discuss optical depth and the generation of random path lengths in standard unpolarised radiative transfer. In Sect.~{\ref{Polarised.sec}} we extend this discussion to polarised radiative transfer, in particular for the case of elongated aligned grains. Apart from the `standard' {\em{extinction}} optical depth, we introduce the {\em{total}} and {\em{dichroic}} optical depth scales, and we compare and analyse them. In Sect.~{\ref{MCRT.sec}} we discuss the implications on these findings for Monte Carlo radiative transfer codes that take dichroic extinction into account. In Sect.~{\ref{DiscussionSummary.sec}} we discuss these results and we sum up.

\section{Unpolarised radiative transfer}
\label{Unpolarised.sec}

As discussed in the Introduction, one of the essential steps in the life cycle of a photon package in a Monte Carlo radiative transfer simulation is the calculation of the next interaction location, or equivalently, the physical path length $s$ covered before the next interaction. To do that, we need to generate a random $s$ from the appropriate probability distribution function $p(s)$. The appropriate PDF $p(s)$ can be found by considering the variation of the specific intensity $I(s)$ along the path. The probability that the photon package has not interacted along the path between $0$ and $s$ is equal to $I(s)/I_0$, with $I_0$ the specific intensity of the photon package at the start of the path.\footnote{Many of the quantities in this paper are dependent on wavelength, including the extinction cross section and the optical depth. In order not to overload the notations, we do not explicitly mention the wavelength dependence.}  Therefore, the {\em{cumulative}} density function $P(s)$ can be written as 
\begin{equation}
P(s) = \int_0^s p(s')\,\txd s' = 1 - \frac{I(s)}{I_0}
\label{P(s)gen}
\end{equation}
In general, we {\em{define}} the optical depth $\tau(s)$ as
\begin{equation}
\tau(s) = -\ln\left[\frac{I(s)}{I_0}\right]
\label{tau}
\end{equation}
Combining these two equations, we find
\begin{equation}
P(s) = 1-e^{-\tau(s)}
\end{equation}
and when we take the derivative of this cumulative density function, 
\begin{equation}
p(\tau) = e^{-\tau}
\end{equation} 
This yields the well-know result that the PDF describing the next interaction location is an exponential distribution in optical depth space. Generating a random optical depth can easily be done by picking a uniform deviate $\calX$, setting $\tau = -\ln(1-\calX)$, and subsequently converting this random $\tau$ to a physical path length $s$. 

The difficulty is that we need to know the solution $I(s)$ to calculate the optical depth scale. This solution is found by solving the appropriate radiative transfer equation. For a photon package moving through an attenuating medium with density $n(s)$ and extinction cross section $C_{\text{ext}}(s)$, the radiative transfer equation reads
\begin{equation}
\frac{\txd I}{\txd s}(s) = -n(s)\,C_{\text{ext}}(s)\,I(s)
\label{RTsimple}
\end{equation}
Note that no additional emission along the path or scattering into the line-of-sight are included, as this is not relevant for this particular photon package. This equation is a simple first-order ordinary differential equation that is easy to solve,
\begin{equation}
I(s) = I_0\,e^{-\tau_{\text{ext}}(s)}
\label{RTsimplesol}
\end{equation}
with the extinction optical depth $\tau_{\text{ext}}(s)$ defined as
\begin{equation}
\tau_{\text{ext}}(s) = \int_0^s n(s')\,C_{\text{ext}}(s')\,\txd s' 
\label{tauext}
\end{equation}
Comparing equations~(\ref{tau}) and (\ref{RTsimplesol}), we see that $\tau(s) = \tau_{\text{ext}}(s)$. In particular, the optical depth scale from which random path lengths can be sampled only depends on the density and optical properties of the material, and not on the properties of the photon package (in this simple case, the only property of the photon package that could matter is $I_0$). 

While the strategy described above is conceptually very simple, some challenges need to be addressed. In particular, the conversion of extinction optical depth to physical path length is usually not a straightforward inversion, and except for some simple idealised cases, needs to be done numerically. In most Monte Carlo radiative transfer codes, the attenuating medium is subdivided into a large number of individual cells, each with a constant density and uniform properties. The codes are typically equipped with a routine to calculate paths through this tessellated medium; this routine returns an ordered list of all the cells $m$ that the path crosses, as well as the length $\Delta s_m$ of the path segments corresponding to the $m$'th cell. Given this ordered list, we can calculate the running values for the path length $s_m$ and the optical depth $\tau_{\text{ext},m}$ at the exit point of each cell crossed by the path. The problem hence reduces to finding the first cell in the array for which $\tau_{\text{ext},m}$ exceeds the randomly generated value of $\tau_{\text{ext}}$, and subsequently applying a linear interpolation to convert $\tau_{\text{ext}}$ to $s$.\footnote{The MC3D radiative transfer code \citep{2008ipid.book.....K, 2012ApJ...751...27H} adopts an alternative method to find the next interaction point. For every dust cell along a path, they generate a new optical depth $\tau$ from an exponential distribution, and they compare this to the extinction optical depth $\Delta\tau_{\text{ext},m}$ within that cell. If $\tau<\Delta\tau_{\text{ext},m}$, the interaction position is determined by linear interpolation; otherwise, the dust cell is crossed and the procedure is repeated for the next cell along the path. This methodology is equivalent to the method used by most other Monte Carlo codes, but seems computationally more expensive, and not straightforward to combine with optimisation techniques as path length stretching \citep{Levitt1968, Spanier1970, 2016A&A...590A..55B} or forced scattering \citep{Cashwell1959, 2013ARA&A..51...63S}.}

These integrations through the dust grid often form the most time-consuming part of a radiative transfer simulation. In order to make these calculations as efficient as possible, especially in 3D geometries, advanced grid construction and grid traversal techniques are required \citep{2006A&A...456....1N, 2008A&A...490..461B, 2012A&A...544A..52L, 2013A&A...560A..35C, 2013A&A...554A..10S, 2014A&A...561A..77S, 2016MNRAS.456..756H}.

\section{Polarised radiative transfer}
\label{Polarised.sec}

\subsection{The Stokes formalism}

The characterisation of the radiation field by the specific intensity, and the corresponding radiative transfer equation (\ref{RTsimple}), are no longer suitable when polarisation is considered. In order to take into account the polarisation state of radiation, one can use the Stokes formalism, which characterises the radiation field by means of the 4D Stokes vector
\begin{equation}
\bfI = \begin{pmatrix} I \\ Q \\ U \\ V \end{pmatrix}
\end{equation}
The first Stokes parameter, $I$,  is still the specific intensity. The Stokes parameters $Q$ and $U$ describe the state of linear polarisation and $V$ describes the state of circular polarisation of the radiation. The Stokes parameters are always defined with respect to a reference direction to be chosen freely from the plane perpendicular to the propagation direction. For a detailed description of the Stokes vector and its connection to the monochromatic transverse electromagnetic waves, we refer to \citet{2000lsnp.book.....M}. 

When we consider the full Stokes vector, the simple radiative transfer equation~(\ref{RTsimple}) becomes
\begin{equation}
\frac{\txd\bfI}{\txd s}(s) = -n(s)\,\bfK(s)\,\bfI(s)
\label{RTE}
\end{equation}
where $\bfK$ is now the extinction matrix, a $4\times4$ matrix that describes how the different Stokes components are affected when radiation passes through the medium. 

\subsection{Spherical grains}

When the dust grains are spherical, or non-spherical but arbitrarily oriented, the extinction matrix is a simple diagonal matrix and all components of the Stokes vector are affected in the same way, and 
\begin{equation}
\bfK = \begin{pmatrix} 
C_{\text{ext}} & 0 & 0 & 0 \\
0 & C_{\text{ext}} & 0 & 0 \\
0 & 0 & C_{\text{ext}} & 0 \\
0 & 0 & 0 & C_{\text{ext}} 
\label{Kspherical}
\end{pmatrix}
\end{equation}
This implies that there is no mixture of the different Stokes components due to extinction, and the solution of the radiative transfer equation can directly be written as
\begin{equation}
\bfI(s) = \bfI_0\,e^{-\tau_{\text{ext}}(s)}
\end{equation}
In particular, the specific intensity $I(s)$ still behaves according to equation (\ref{RTsimplesol}), exactly as in the case of non-polarised radiative transfer. We can immediately conclude that, also in this case, the PDF describing the net interaction location is an exponential distribution in extinction optical depth space. 

\subsection{Spheroidal grains}
\label{SpheroidalGrains.sec}

Complexity arises when the dust grains are non-spherical and (partially) aligned. In this case, the extinction matrix $\bfK$ is not a diagonal matrix, but a full $4\times4$ matrix with 16 nonzero cross sections, but only seven independent ones \citep{1957lssp.book.....V, 1996JQSRT..55..649H}. Fortunately, in the case of spheroidal grains, $\bfK$ is significantly less complex. If we denote the orientation of the grain alignment at a distance $s$ along the path as $\psi(s)$, $\bfK$ can be expressed as \citep{1974ApJ...187..461M, 2002A&A...385..365W}
\begin{equation}
\bfK(s) = {\mathcal{L}}\Bigl(-\psi(s)\Bigr)\,\bfK_{\text{ref}}(s)\,{\mathcal{L}}\Bigl(\psi(s)\Bigr)
\label{Kfull}
\end{equation}
with ${\mathcal{L}}(\psi)$ a Mueller rotation matrix
\begin{equation}
{\mathcal{L}}(\psi) =
\begin{pmatrix} 
1 & 0 & 0 & 0 \\
0 & \cos2\psi & \sin2\psi & 0 \\
0 & -\sin2\psi & \cos2\psi & 0 \\
0 & 0 & 0 & 1
\end{pmatrix}
\end{equation}
and $\bfK_{\text{ref}}$ the extinction matrix in the reference frame of the grain,
\begin{equation}
\bfK_{\text{ref}} = \begin{pmatrix} 
C_{\text{ext}} & C_{\text{pol}} & 0 & 0 \\
C_{\text{pol}} & C_{\text{ext}} & 0 & 0 \\
0 & 0 & C_{\text{ext}} & C_{\text{cpol}} \\
0 & 0 & -C_{\text{cpol}} & C_{\text{ext}} 
\end{pmatrix}
\label{Ksph}
\end{equation}
Note that $\bfK_{\text{ref}}$ has only three independent elements: the extinction cross section $C_{\text{ext}}$, polarisation cross section $C_{\text{pol}}$, and circular polarisation cross section $C_{\text{cpol}}$ \citep{2000lsnp.book.....M, 2002ApJ...574..205W}. 

The non-diagonal character of the extinction matrix has an important effect on the radiation field: the different components of the Stokes vector are coupled and will get mixed along the path. In particular, radiation that is initially unpolarised can develop linear and even circular polarisation just by propagating through the medium \citep{SERKOWSKI1962289, 1972MNRAS.159..179M, 1974ApJ...187..461M}.

This dichroism complicates the relation between path length and optical depth. To stress the fact that we deal with dichroic extinction, we will use the term dichroic optical depth and use the notation $\tau_{\text{dic}}(s)$ when we consider the optical depth in a dichroic medium. In general we can write that 
\begin{equation}
\tau_{\text{dic}}(s) = -\ln\left[\frac{I(s)}{I_0}\right],
\end{equation}
where $I(s)$ should now be seen as the first component of the Stokes vector $\bfI(s)$, obtained by solving the vector radiative transfer equation (\ref{RTE}) with $\bfK(s)$ given by (\ref{Kfull}).

In the case of spherical grains, we have shown that $\tau(s) = \tau_{\text{ext}}(s)$. This equivalence does not hold for the general case of spheroidal grains, however. Indeed, with an extinction matrix given by expression~(\ref{Kfull}), the radiative transfer equation is a set of four coupled first-order ordinary differential equations with varying coefficients. The full solution for $I(s)$, and hence the dichroic optical depth $\tau_{\text{dic}}(s)$, will depend on all elements of the extinction matrix, and on all initial Stokes components. As the extinction optical depth (\ref{tauext}) only depends on one single element ($K_{11} = C_{\text{ext}}$) of the extinction matrix and is independent of the initial polarisation state $\bfI_0$, it is impossible that the dichroic optical depth is equivalent to the extinction optical depth.

One could consider another optical depth scale that does depend on the initial polarisation state of the photon package. An interesting starting point is the {\em{total}} extinction cross section, 
\begin{equation}
\tilde{C}_{\text{ext}}
=
K_{11} + \frac{Q_0}{I_0}\,K_{12} + \frac{U_0}{I_0}\,K_{13} + \frac{V_0}{I_0}\,K_{14}
\label{deftotalC}
\end{equation}
introduced by \citet{2000lsnp.book.....M} and \citet{2002ApJ...574..205W}. It properly describes the attenuation of an incident beam with arbitrary initial polarisation state $\bfI_0$. Based on this total extinction cross section, one can define an optical depth scale, which we denote as the {\em{total}} optical depth, in a similar way as we had for the extinction optical depth (\ref{tauext}), 
\begin{equation}
\tau_{\text{tot}}(s) = \int_0^s n(s')\,\tilde{C}_{\text{ext}}(s')\, \txd s' 
\label{tautot}
\end{equation}
or explicitly,
\begin{multline}
\tau_{\text{tot}}(s)
=
\tau_{\text{ext}}(s) 
+ \frac{Q_0}{I_0}\int_0^s n(s')\,C_{\text{pol}}(s')\cos2\psi(s')\,\txd s' 
\\+ \frac{U_0}{I_0}\int_0^s n(s')\,C_{\text{pol}}(s')\sin2\psi(s')\,\txd s' 
\label{tautot2}
\end{multline}
In spite of the increased complexity, it is computationally not much more difficult than the extinction optical depth. Indeed, for a photon package in a Monte Carlo simulation, with a given position, propagation direction and polarisation state, it is relatively easy to compute the total optical depth $\tau_{\text{tot},m}$ at the entry point of each cell crossed by the path.

We can, however, make an analogous reasoning as above to argue that the total optical depth cannot be equivalent to the dichroic optical depth. Indeed, with an extinction matrix given by expression~(\ref{Kfull}), the radiative transfer equation is a set of four coupled first-order ordinary differential equations with varying coefficients. The full solution for $I(s)$, and hence the dichroic optical depth $\tau_{\text{dic}}(s)$, depends on {\em{all}} elements of the extinction matrix, and on {\em{all}} initial Stokes components. The expression~(\ref{tautot2}) does depend on the initial linear polarisation state of the incoming radiation and on the polarisation cross section, but it does not involve any dependency on the circular polarisation cross section $C_{\text{cpol}}$ or the initial circular polarisation $V_0$. As a result, it is impossible that the total extinction is fully equivalent to the dichroic optical depth.

If neither the extinction optical depth, nor the total optical depth are equivalent to the dichroic optical depth, what is the correct expression for the dichroic optical depth? For a medium with non-uniform grain alignment, the answer to this question is somewhat disappointing: given the non-trivial coupling of all four components of the radiative transfer equation, it is not possible to write down a closed formula for $I(s)$, nor for the dichroic optical depth $\tau_{\text{dic}}(s)$. Obtaining the solution $I(s)$ can be achieved numerically using standard vector ODE solution techniques. One might wonder whether $\tau_{\text{ext}}$ or $\tau_{\text{tot}}$, for which we have a closed expression, are reasonable approximations for the dichroic optical depth. To answer this question, we have performed some numerical tests. 

We have adopted a hypothetical example where $n(s) = \psi(s) = s$, i.e.\ the density of material increases linearly with increasing distance, and the grain alignment rotates around the path. Furthermore, we have assumed that the optical properties do not vary along the path, and we set $C_{\text{ext}} = 1$ and $C_{\text{pol}} = C_{\text{cpol}} = 0.1$ (representative values of $C_{\text{pol}}/C_{\text{ext}}$ go up to 0.3 and more). On the one hand, we have calculated the dichroic optical depth $\tau(s)$ by numerically solving the vector radiative transfer equation using an explicit Runge-Kutta method with variable step size control. On the other hand, we have used equations~(\ref{tauext}) and (\ref{tautot}) to calculate $\tau_{\text{ext}}(s)$ and $\tau_{\text{tot}}(s)$, and we have calculated the relative differences
\begin{equation}
\delta_{\text{ext}} = \frac{\tau_{\text{ext}}-\tau_{\text{dic}}}{\tau_{\text{dic}}}
\qquad
\delta_{\text{tot}} = \frac{\tau_{\text{tot}}-\tau_{\text{dic}}}{\tau_{\text{dic}}}
\end{equation}
We calculated these quantities for initial Stokes vectors that are 100\% linearly polarised, with the linear polarisation angle 
\begin{equation}
\theta = \frac12\arctan\frac{U_0}{Q_0} = \frac12\arccos\frac{Q_0}{I_0}
\end{equation}
gradually changing between 0$^\circ$ and 90$^\circ$ in steps of 5$^\circ$. 

\begin{figure*}
\includegraphics[width=0.95\textwidth]{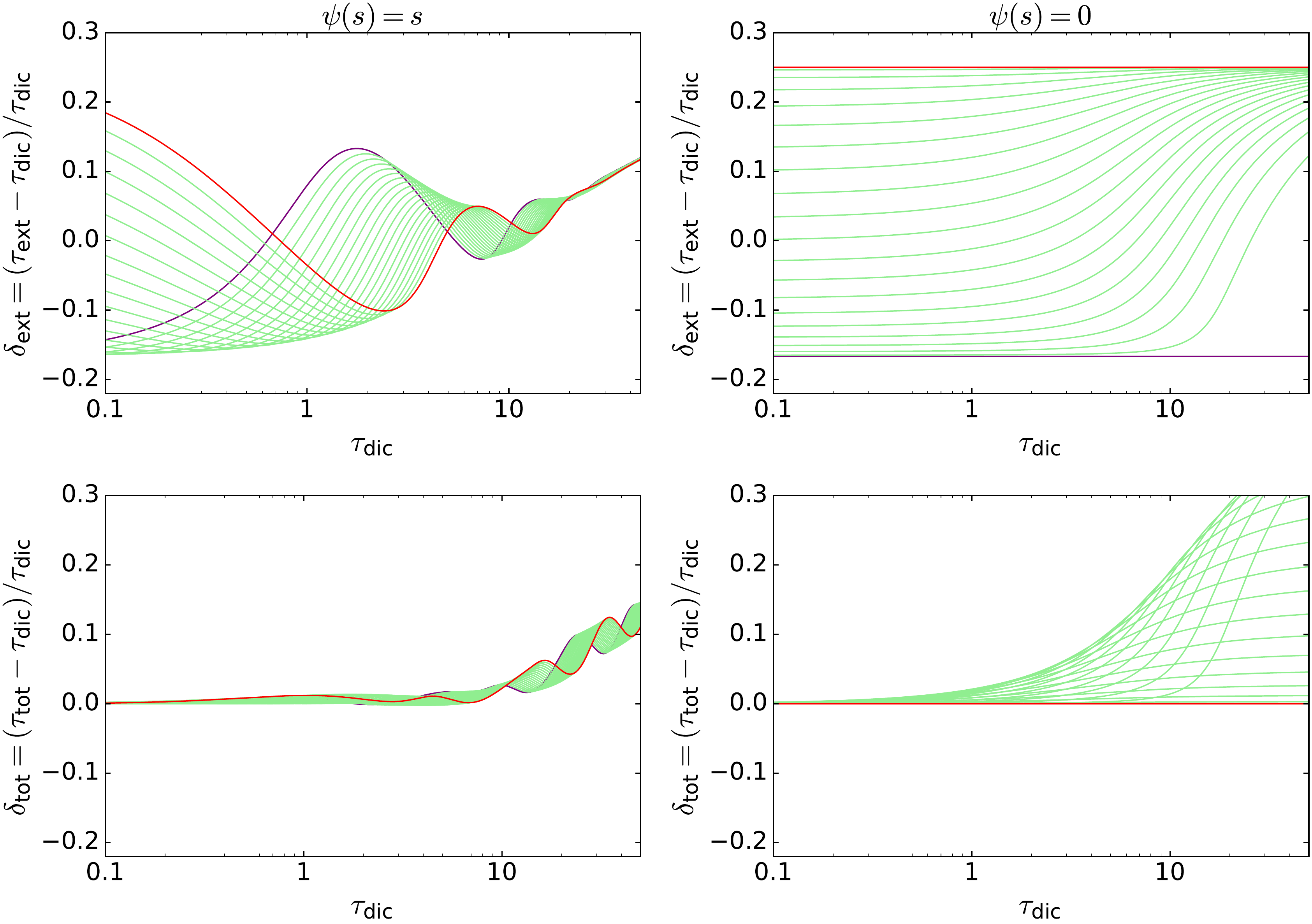}
\caption{Top left: the relative differences $\delta_{\text{ext}}$ between the extinction optical depth $\tau_{\text{ext}}$ and the dichroic optical depth $\tau_{\text{dic}}$ for the hypothetical example with density and grain alignment increasing linearly with increasing path length. The different curves correspond to different input Stokes vectors: in each case the photon package is initially fully linearly polarised, but the Stokes vector orientation $\theta$ increases in steps of 5$^\circ$. The red lines corresponds to $\theta=0^\circ$ ($Q_0/I_0=1$), the purple line to $\theta=90^\circ$ ($Q_0/I_0=-1$). Bottom left: similar as top left panel, but now showing the relative difference $\delta_{\text{tot}}$ between the total optical depth $\tau_{\text{tot}}$ and the dichroic optical depth. Right: same as the panels on the left, but now for a model with uniform grain alignment.}
\label{Alignment.fig}
\end{figure*}

In the upper left panel of Figure~{\ref{Alignment.fig}} we show $\delta_{\text{ext}}$ as a function of the dichroic optical depth $\tau_{\text{dic}}$ along the path. The extinction optical depth scale has the advantage that it is simple and does not explicitly depend on the polarisation state of the photon package, such that the relation between $s$ and $\tau_{\text{ext}}$ can in principle be precomputed. It is clear, however, that $\tau_{\text{ext}}$ is a poor approximation for the dichroic optical depth: depending on the value of $Q_0/I_0$ it can both underestimate and overestimate $\tau_{\text{dic}}$ by up to 20 percent. Most importantly, the differences between $\tau_{\text{dic}}$ and $\tau_{\text{ext}}$ can be large even at small optical depths. 

The relative difference $\delta_{\text{tot}}$ in the bottom left panel shows a very different behaviour. The total optical depth always overestimates the dichroic optical depth. A second important difference is that $\delta_{\text{tot}}$ is usually smaller than (the absolute value of) $\delta_{\text{ext}}$. This is particularly true in the optically thin limit, where $\tau_{\text{tot}}$ approximates $\tau_{\text{dic}}$ very well.  At large optical depths, $\tau_{\text{dic}}>10$, it turns out that the total optical depth is not always a reliable approximation to the dichroic optical depth: $\tau_{\text{tot}}$ can even become a poorer approximation to the dichroic optical depth than the simpler approximation $\tau_{\text{ext}}$. 

\subsection{Spheroidal grains with uniform alignment}

We can gain more insight into these results by considering the special case $\psi(s)=0$, i.e., the grains are all uniformly aligned along the path.\footnote{We can always perform a rotation to the initial Stokes vector to ensure that it is aligned with the grain orientation.} In this special case, the rotation matrices in (\ref{Kfull}) are the identity matrices, and $\bfK=\bfK_{\text{ref}}$. With this relatively simple block-diagonal extinction matrix, the four different components of the Stokes vector are paired instead of fully coupled: $I$ and $Q$ are linked, and $U$ and $V$. It is possible write down the full solution of the radiative transfer equation,
\begin{subequations}
\label{solRTE-al}
\begin{align}
I(s) &= e^{-\tau_{\text{ext}}(s)} \left[I_0\cosh\tau_{\text{pol}}(s) - Q_0\sinh\tau_{\text{pol}}(s)\right] 
\label{I(s)gen} \\
Q(s) &= e^{-\tau_{\text{ext}}(s)} \left[Q_0\cosh\tau_{\text{pol}}(s) - I_0\sinh\tau_{\text{pol}}(s)\right] \\
U(s) &= e^{-\tau_{\text{ext}}(s)} \left[U_0\cos\tau_{\text{cpol}}(s) - V_0\sin\tau_{\text{cpol}}(s)\right] \\
V(s) &= e^{-\tau_{\text{ext}}(s)} \left[V_0\cos\tau_{\text{cpol}}(s) + U_0\sin\tau_{\text{cpol}}(s)\right]
\end{align}
\end{subequations}
with
\begin{align}
\tau_{\text{pol}}(s) &= \int_0^s n(s')\,C_{\text{pol}}(s')\,\txd s' \\
\tau_{\text{cpol}}(s) &= \int_0^s n(s')\,C_{\text{cpol}}(s')\,\txd s'
\end{align}
Combining equation (\ref{tau}) with the solution (\ref{I(s)gen}) for the specific intensity, we find an explicit expression for the dichroic optical depth 
\begin{equation}
\tau_{\text{dic}}(s) = \tau_{\text{ext}}(s)
-\ln\left[\cosh\tau_{\text{pol}}(s) - \frac{Q_0}{I_0}\sinh\tau_{\text{pol}}(s)\right]
\label{tau-sph}
\end{equation}
In the right-hand side panels of Figure~{\ref{Alignment.fig}} we show $\delta_{\text{ext}}$ and $\delta_{\text{tot}}$ as a function of $\tau_{\text{dic}}$ along the path for the case with $\psi=0$. Again, $\tau_{\text{ext}}$ is a poor approximation for the dichroic optical depth, even at small optical depths. Based on the explicit expression~(\ref{tau-sph}), we can calculate the extreme values for $\delta_{\text{ext}}$, which correspond to $Q_0/I_0 = \pm1$, 
\begin{equation}
\delta_{\text{ext}}^{\pm} = \mp\frac{C_{\text{pol}}}{C_{\text{ext}} \pm C_{\text{pol}}}
\end{equation} 
For our example, these differences run up to $-17$ and 25\%. On the other hand, the total optical depth is in general a better approximation to the dichroic optical depth, especially in the optically thin limit $\tau\ll1$. This can be understood by considering the Taylor expansion for expression (\ref{tau-sph}) for $\tau_{\text{pol}}\ll1$, 
\begin{equation}
\tau_{\text{dic}} = 
\underbrace{\tau_{\text{ext}} 
+ \frac{Q_0}{I_0}\,\tau_{\text{pol}}}_{\tau_{\text{tot}}}
- \frac12\left[1-\left(\frac{Q_0}{I_0}\right)^2\right]\tau_{\text{pol}}^2 
+ {\mathcal{O}}\Bigl(\tau_{\text{pol}}^3\Bigr)
\end{equation}
This result shows that the extinction optical depth approximates the dichroic optical depth to first order in $\tau_{\text{pol}}$, whereas the total optical depth approximates it to second order. For small optical depths, it is hence not surprising that the dichroic optical depth is a much better approximation. Moreover, as the coefficient of the second order term is always negative, the total optical depth systematically overestimates the dichroic optical depth. Note, however, that the total optical depth is not guaranteed to be a reliable estimator for the dichroic optical depth: at large optical depths, $\tau_{\text{tot}}$ overestimates $\tau_{\text{dic}}$ significantly.

\section{Application in Monte Carlo radiative transfer}
\label{MCRT.sec}

\subsection{Random generation of path lengths}
\label{Random.sec}

In the previous Section, we have demonstrated that neither the simple extinction optical depth, nor the total optical depth are equivalent to the dichroic optical depth. In particular, we have seen that, for a given physical path length, relative differences of up to several ten percent are possible between these optical depth scales (for the total optical depth scale, these large differences only occur at large optical depths). As discussed in the Introduction, the optical depth in Monte Carlo radiative transfer simulations is mainly important for the random generation of the next interaction location. In practice, a random optical depth is generated from an exponential PDF, and this optical depth is translated to a physical path length, which immediately sets the next interaction location. 

\begin{figure*}
\includegraphics[width=0.98\textwidth]{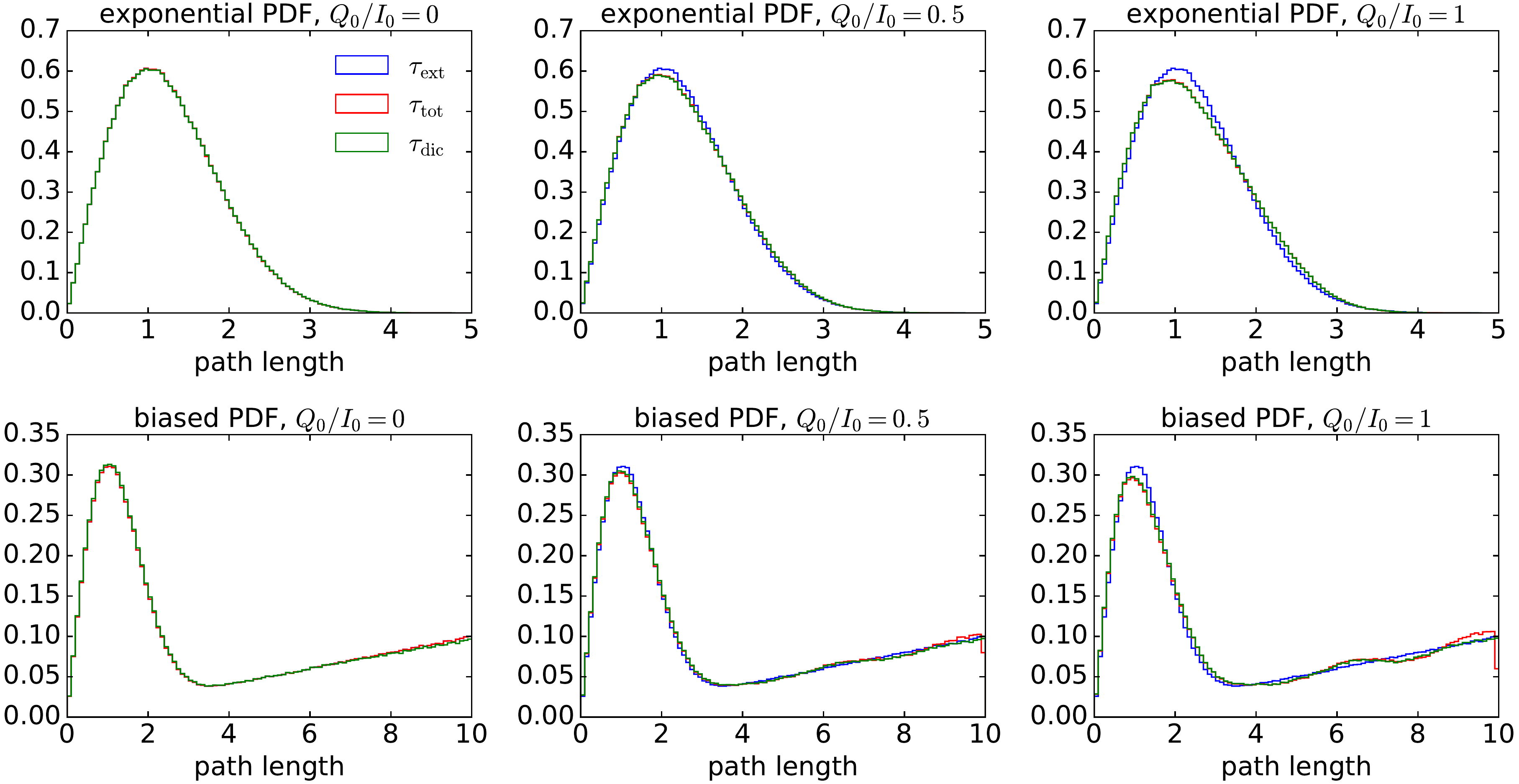}
\caption{Histograms for the distribution of randomly generated path lengths, corresponding to the different optical depth scales discussed in this paper. The model is described in Sect.~{\ref{Random.sec}}, and the different columns correspond to different levels of initial linear polarisation. The panels on the top correspond to optical depths randomly generated from an exponential distribution, the panels on the bottom row to optical depths generated using the composite biasing approach discussed in \citet{2016A&A...590A..55B}.}
\label{FindPathLength.fig}
\end{figure*}

In order to find out to which degree the choice of the optical depth scale affects the determination of the path length, we performed a number of simple simulations. We adopted a similar setup as in Sect.~{\ref{SpheroidalGrains.sec}}, with $n(s)=\psi(s)=s$, $C_{\text{ext}}=1$, $C_{\text{pol}}=C_{\text{cpol}}=0.1$, and we used a maximum path length $s_{\text{max}}=10$, corresponding to $\tau_{\text{ext},{\text{max}}} = 50$. We generated one million random optical depths from an exponential PDF, and we converted these values to physical path lengths according to each of the three different optical depth scales. 

The top row of Fig.~{\ref{FindPathLength.fig}} shows the histograms of the corresponding path lengths, for three different input linear polarisations ($Q_0/I_0=0$, $0.5$ and $1$, from left to right). For initially unpolarised photon packages, the three distributions are very similar (in fact, the distributions corresponding the extinction and total optical depth are identical, as $\tau_{\text{ext}}(s)\equiv\tau_{\text{tot}}(s)$ for initially unpolarised radiation). When the initial linear polarisation degree increases, the histograms for the dichroic and total optical depth scales remain very similar, but they gradually deviate more from the histogram corresponding to the extinction optical depth. We applied two-sample Kolmogorov-Smirnov tests to quantify this observation. These tests demonstrated that, indeed, there is no significant difference between the path length distribution corresponding to the total optical depth scale and the dichroic optical depth scale, whereas the path length distribution corresponding to the extinction optical depth scale is significantly different (in the case of initially linearly polarised photon packages). This result is not surprising, as the exponential PDF strongly favours small random optical depths, and in the optically thin regime, the total optical depth approximates the dichroic optical depth very well (see Fig.~{\ref{Alignment.fig}}).

It is not always advisable to randomly generate optical depths from an exponential distribution, though. One particularly relevant case is the penetration of radiation through an optically thick medium, which is a notoriously difficult task for Monte Carlo radiative transfer \citep[e.g.,][]{2009A&A...497..155M, 2017A&A...603A.114G, 2018ApJ...861...80C}. For a standard exponential PDF, huge numbers of photon packages need to be generated before a single one might cross the barrier. Path length stretching \citep{Levitt1968, Spanier1970, Dwivedi1982} is a Monte Carlo optimisation technique that artificially stimulates the generation of larger path lengths. \citet{2016A&A...590A..55B} combined this technique with the concept of composite biasing, which results in an approach that has the advantages of path length stretching (an increased probability for large path lengths) and minimises the disadvantages (no large weight factors for the photon packages). The bottom row of Fig.~{\ref{FindPathLength.fig}} shows the same histograms as the top row, but now the random optical depths are sampled from a biased distribution \citep[for details, see][Sect.~4.3]{2016A&A...590A..55B}. As the total optical depth is no longer guaranteed to be a good approximation to the dichroic optical depth at high optical depths, it is no surprise that the distributions are now more different, especially in the high path length tail. KS tests indicate that the probability that they are drawn from the same distribution is negligible.

\subsection{Practical calculation of the dichroic optical depth}

The results of the previous subsections show that it is not a good idea to use the extinction or total optical depth scale to generate random path lengths in a Monte Carlo radiative transfer simulation with non-spherical dust grains. To avoid systematic errors, the dichroic optical depth scale should be used to translate randomly generated optical depths to physical path lengths. We hence need an efficient calculation of the dichroic optical depth $\tau_{\text{dic}}(s)$ along the path. However, in Sect.~{\ref{SpheroidalGrains.sec}} we argued that it is not possible to write down a general closed formula for $\tau_{\text{dic}}(s)$, similar to the ``simple'' formulae (\ref{tauext}) and (\ref{tautot2}) for the extinction and total optical depth scales, respectively. In general, a solution for $\tau_{\text{dic}}(s)$ needs be obtained numerically using standard vector ODE solution techniques. 

Fortunately, in the context of Monte Carlo radiative transfer simulations, where the density of the medium is usually discretised on a grid with a uniform density and uniform optical properties within each grid cell, there is an efficient routine to solve the radiative transfer equation without the need to use vector ODE solution methods \citep[see also][]{2002ApJ...574..205W, 2003JQSRT..79..921L}. Indeed, we can progressively solve the radiative transfer equation in each individual cell along the path. Assume again that the path is split into individual cells, small enough that the density, optical properties and grain orientation can be considered uniform within each cell, and denote the Stokes vector at the entry point of cell $m$ as $\bfI_{m-1}$. We first rotate the Stokes vector over an angle $\psi_m$ to align it with the grain orientation within cell $m$,
\begin{equation}
\bfI_{m-1}^\prime = {\mathcal{L}}(\psi_m)\,\bfI_{m-1}
\end{equation}
As the grain orientation with the cell is constant, we can directly apply the solution (\ref{solRTE-al}) for the radiative transfer to calculate the Stokes vector at the exit point $s=s_m$ of the cell,
\begin{gather}
\Delta\tau_{\text{xxx},m} = n_m\,C_{\text{xxx},m}\,\Delta s_m \\
I_m^\prime = e^{-\Delta\tau_{\text{ext},m}} 
\left[I_{m-1}^\prime \cosh\Delta\tau_{\text{pol},m} - Q_{m-1}^\prime\sinh\Delta\tau_{\text{pol},m}\right] 
\\
Q_m^\prime = e^{-\Delta\tau_{\text{ext},m}} 
\left[Q_{m-1}^\prime \cosh\Delta\tau_{\text{pol},m} - I_{m-1}^\prime\sinh\Delta\tau_{\text{pol},m}\right] 
\\
U_m^\prime = e^{-\Delta\tau_{\text{ext},m}} 
\left[U_{m-1}^\prime \cos\Delta\tau_{\text{cpol},m} - V_{m-1}^\prime\sin\Delta\tau_{\text{cpol},m}\right] 
\\
V_m^\prime = e^{-\Delta\tau_{\text{ext},m}} 
\left[V_{m-1}^\prime \cos\Delta\tau_{\text{cpol},m} + U_{m-1}^\prime\sin\Delta\tau_{\text{cpol},m}\right] 
\end{gather}
When we rotate the resulting Stokes vector back to the laboratory frame, we find the desired result $\bfI_m$,
\begin{equation}
\bfI_m = {\mathcal{L}}(-\psi_m)\,\bfI_m^\prime
\end{equation}
This recipe can be repeated for all cells along the path.

\begin{figure}
\includegraphics[width=0.45\textwidth]{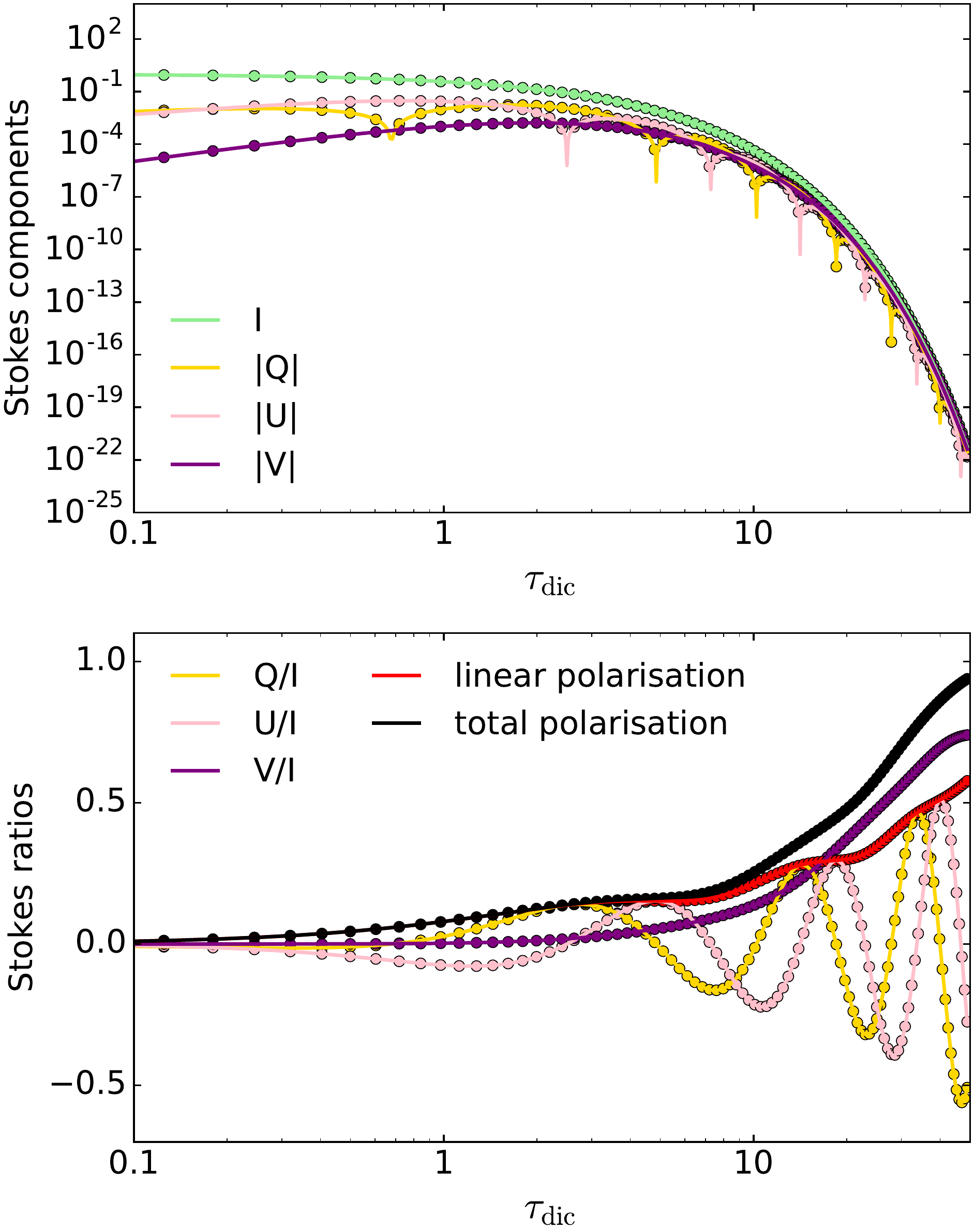}
\caption{Evolution of an initially unpolarised Stokes vector propagating through a medium of aligned grains where the grain alignment rotates around the path. The dots are the result from a Runge-Kutta numerical integration of the vector radiative transfer equation, the solid lines correspond to the method outlined in Sect.~{\ref{Polarised.sec}}.}  
\label{RotatingGrains.fig}
\end{figure}

We have tested this strategy using the same example as discussed before. The comparison between the brute-force Runge-Kutta approach and the algorithm described above is shown in Figure~{\ref{RotatingGrains.fig}} for an initially unpolarised Stokes vector. The top panel shows the evolution of the individual Stokes components, the bottom panels shows the Stokes ratios, as well as the degree of linear and total polarisation,
\begin{equation}
p_{\text{L}} = \frac{\sqrt{Q^2+U^2}}{I},\quad
p = \frac{\sqrt{Q^2+U^2+V^2}}{I}
\end{equation}
The two methods clearly agree. As the photon, initially unpolarised, propagates along the path, it gradually develops linear polarisation as a result of dichroic extinction, and later on also circular polarisation. From $\tau\gtrsim8$, the circular polarisation starts to dominate the linear polarisation, and at $\tau=50$, the photon is almost 100\% polarised.

The determination of a random path length now follows the same strategy as discussed for the unpolarised Monte Carlo radiative transfer in Sect.~{\ref{Unpolarised.sec}}. From the solution of the Stokes vector $\bfI_m$, we calculate the dichroic optical depth $\tau_{\text{dic},m}$ at the exit point of each cell, and we search for the first cell for which $\tau_{\text{dic},m}$ exceeds the randomly determined $\tau$. One additional difference needs to be taken into account. In the case of unpolarised radiation transfer, the increase in optical depth {\em{within}} each cell is directly proportional to the increase in path length within that cell,
\begin{equation}
\tau_{\text{ext}}(s) = \tau_{\text{ext},m-1}+n_m\,C_{\text{ext},m}\,(s-s_{m-1})
\qquad{s_{m-1}\leq s\leq s_m}
\end{equation}
To find the exact path length $s$ corresponding to a randomly generated $\tau$, we can therefore use simple linear interpolation,
\begin{equation}
s -s_{m-1} = \frac{\tau-\tau_{\text{ext},m-1}}{n_m\,C_{\text{ext},m}} = \left(\frac{s_m-s_{m-1}}{\tau_{\text{ext},m}-\tau_{\text{ext},m-1}}\right)(\tau-\tau_{\text{ext},m-1})  
\end{equation}
In the case of dichroic attenuation, the increase in optical depth within a cell is no longer proportional to the increase in path length, as
\begin{align}
\tau_{\text{dic}}(s) = \tau_{\text{dic},m-1}&+n_m\,C_{\text{ext},m}\,(s-s_{m-1}) \nonumber \\
&-\ln\,\Biggl[ \cosh\, \Bigl(n_m\,C_{\text{pol},m}\,(s-s_{m-1})\Bigr) \nonumber \\
&\qquad\quad- \frac{Q_{m-1}^\prime}{I_{m-1}^\prime}
\sinh \,\Bigl(n_m\,C_{\text{pol},m}\,(s-s_{m-1})\Bigr) \Biggr] \nonumber \\
&\hspace{12em}{s_{m-1}\leq s\leq s_m}
\end{align}
For a randomly generated $\tau$, we should in principle use this equation to determine the correct value of $s$, which can be done using standard root-finding algorithms. In practice, however, we suggest to use linear interpolation: given the approximations due to the discretisation itself, the gain in accuracy by applying an exact root finding algorithm is probably not worth the additional computational cost. Only in cases where the individual cells have a high optical depth, it could possibly be useful to consider a more advanced (and numerically more costly) higher-order interpolation scheme.

One could argue that, in spite of the errors made, it would still be advantageous to use the extinction or total optical depth instead of the dichroic optical depth, because the calculation of the dichroic optical depth is numerically more demanding. Indeed, calculating $\tau_{\text{ext}}(s)$ or $\tau_{\text{tot}}(s)$ involves just a single summation along the path, whereas the calculation of $\tau_{\text{dic}}(s)$ requires the propagation of the entire Stokes vector, including rotations and hyperbolic function evaluations at every grid cell. However, it should be realised that this operation has to be executed anyway in the Monte Carlo loop: the calculation of the Stokes vector up to the next interaction point is required, because the albedo and the scattering matrix explicitly depend on the polarisation state of the radiation \citep{2000lsnp.book.....M, 2002A&A...385..365W}. Rather than being a numerically expensive extra, the calculation of the dichroic optical depth comes for free. We can conclude that there is no benefit at all in using $\tau_{\text{ext}}$, $\tau_{\text{tot}}$, or any other approximation, instead of the dichroic optical depth to calculate the next interaction location.

\section{Summary and outlook}
\label{DiscussionSummary.sec}

We have performed an analysis of the attenuation of radiation when it passes through a medium of aligned spheroidal grains, fully taking into account the effects of dichroism. The most important conclusions from this analysis are the following:
\begin{itemize}
\item In a dichroic medium, the dichroic optical depth is no longer equivalent to the usual extinction optical depth $\tau_{\text{ext}}$, i.e.\ the integral of the product of number density and extinction cross section along the path. For representative values of the optical properties of dust grains, the relative difference between both optical depth scales can be several ten percent, even at low optical depths.
\item A potential option to account for dichroic attenuation could be to replace the extinction cross section by the total extinction cross section. The corresponding total optical depth $\tau_{\text{tot}}$ approximates the dichroic optical depth to first order, but always overestimates it. Relative differences between total and dichroic optical depth are small at low optical depths, but can also run up to several ten percent at high optical depths. 
\item An accurate calculation of the dichroic optical depth requires the full solution of the intensity profile along the path. In the general case of a dichroic medium, the radiative transfer equation becomes a set of four coupled first-order differential equations with varying coefficients, and a closed expression for the dichroic optical depth cannot be derived. However, the exact solution corresponding to a medium with uniform grain alignment can be used to find the full solution in an elegant way without any further numerical integration. There is no benefit in using $\tau_{\text{ext}}$, $\tau_{\text{tot}}$, or any other approximation instead of the dichroic optical depth to calculate the next interaction location in a Monte Carlo radiative transfer simulation.
\end{itemize}
Our results have implications for Monte Carlo radiative transfer codes that wish to incorporate the attenuation by elongated dust grains. If scattering polarisation by spherical grains already adds some complexity to Monte Carlo radiative transfer codes, dealing with non-spherical grains increases this complexity to a new level. Compared to spherical grains, the scattering process is significantly more complex. The scattering properties of spherical grains are fully described by just the albedo and the scattering phase function; for elongated grains, a full $4\times4$ scattering matrix comes into play \citep{2000lsnp.book.....M}, and the random determination of a new propagation direction after a scattering event is not trivial \citep{2002A&A...385..365W, 2002ApJ...574..205W, 2003JQSRT..79..921L}. A related complexity concerns the amount of data that needs to be stored and accessed: each of the elements of the extinction matrix and the scattering matrix is not only dependent on grain material, size and wavelength, but also on shape and incidence angle. Moreover, each element of the scattering matrix needs to be discretised on the unit sphere. Finally, the process of dichroic extinction adds yet another level of complexity, and the results in this paper show that this also affects the random generation of the next interaction location.

So far, only a limited number of Monte Carlo radiative transfer codes have attempted to actually calculate dichroic attenuation by non-spherical aligned grains.\footnote{Several codes \citep{1997ApJ...477L..25W, 1997AJ....114.1405W, 2018arXiv180606101S} use an approximate treatment of dichroic attenuation, based on a nonlinear relationship between the magnitude of dichroic polarisation and optical depth in our Milky Way \citep{1989ApJ...346..728J, 2008ApJ...674..304W}.} \citet{2002A&A...385..365W} were the first to include non-spherical aligned grains in their Monte Carlo radiative transfer calculations. They presented multiple light scattering calculations, and demonstrated that the incorporation of elongated grains is important to explain the circular polarisation of light. They discussed the concepts of dichroism and birefringence, but did not include these effects in the simulations they presented. \citet{2002ApJ...574..205W} presented a Monte Carlo code that models the effects of scattering and dichroic absorption by aligned grains in circumstellar environments. While their code is presented for general geometries, they only discussed models with a uniform grain alignment. \citet{2013MNRAS.435.3419S} applied this code to massive young stellar objects with more complex magnetic field configurations. \citet{2003JQSRT..79..921L} presented a third Monte Carlo code with more or less the same characteristics as the code by \citet{2002ApJ...574..205W}, and also with the modelling of young stellar objects as the prime science objective. Applications of this code were presented by \citet{2004MNRAS.352.1347L} and \citet{2007Natur.450...71C}. Peest et al.\ (in prep.) discuss the implementation of polarisation by elongated grains in the vectorised Monte Carlo code MC3D developed by \citep{2008ipid.book.....K, 2012ApJ...751...27H}. Finally, a new Monte Carlo radiative transfer code, POLARIS, was presented by \citet{2016A&A...593A..87R}. It handles dichroic extinction and polarised emission, and is optimised to handle data that results from sophisticated magneto-hydrodynamic simulations \citep{2017A&A...601A..90B, 2017A&A...603A..71R, 2018A&A...611A..70R}. 

The work we have presented here fits into a broader effort to fully integrate the attenuation, polarisation and thermal emission by elongated interstellar dust grains into the publicly available radiative transfer code SKIRT \citep{2015A&C.....9...20C}. The advantage of implementing elongated grains in SKIRT is that the code can then use many of the useful ingredients that are already available, such as a suite of optimisation techniques \citep{2011ApJS..196...22B, 2016A&A...590A..55B}, a library of input geometries for sources and sinks \citep{2015A&C....12...33B}, advanced spatial grids and grid traversal techniques \citep{2013A&A...560A..35C, 2013A&A...554A..10S, 2014A&A...561A..77S}, the coupling to the output of grid-based and particle-based hydrodynamic codes \citep{2015A&A...576A..31S, 2016MNRAS.462.1057C}, and hybrid parallelisation techniques for shared and distributed memory machines \citep{2017A&C....20...16V}. 

Contrary to the currently available radiative transfer codes that incorporate elongated grains, SKIRT mainly focuses on galaxy-wide scales \citep[e.g.,][]{2012MNRAS.427.2797D, 2014A&A...571A..69D, 2017A&A...599A..64V, 2017MNRAS.470..771T}. The magnetic and turbulent energy densities in nearby galaxies are found to be roughly in equipartition, and therefore magnetic fields are expected to be important for the evolution of galaxies \citep{1990ApJ...365..544B, 1996ARA&A..34..155B}. High-resolution cosmological zoom simulations have recently started to take into account magnetic fields \citep{2014ApJ...783L..20P, 2017MNRAS.469.3185P, 2017MNRAS.467..179G}. Radiative transfer codes that can fully incorporate dichroic extinction and emission by elongated aligned grains could be important tools to compare such simulations to observations.

\bibliographystyle{aa}
\bibliography{OpticalDepth}

\end{document}